\documentclass{article} 
\usepackage{amsmath,amssymb,graphicx}
\pdfoutput=1 

\begin{document}

\title{On the upper limits for dipole anisotropies in cosmic-ray positrons}
\author{J.~Berdugo, J.~Casaus, C.~Mana, M.A.~Velasco\\
\small Centro de Investigaciones Energ\'eticas, Medioambientales y Tecnol\'ogicas, CIEMAT.\\
\small Av. Complutense 40, Madrid E-28040, Spain\\
}
\date{}
\maketitle

\normalsize

\begin{abstract}
The excess of cosmic-ray positrons in the energy range from 10 GeV to few hundred GeV
reported by PAMELA and AMS experiments is not consistent with a pure secondary origin
and requires the introduction of a source term. The presence of anisotropies in the
positron arrival directions would be a distinctive signature of their origin. Current
measurements are consistent with isotropy and limits to a dipole anisotropy have been
established. In this note, we review the mathematical basis of this analysis and provide
a general bound to the dipole upper limits achievable from a given sample of events.
The published experimental limits are confronted with this bound.
\end{abstract}

\section{Introduction}

Recent measurements of the cosmic-ray positrons carried out by the
PAMELA~\cite{ref:pamela-ep} and AMS~\cite{ref:ams-ep}
space spectrometers have provided precise characterization of their energy spectrum up
to few hundred GeV. In particular, the positron spectrum above $\sim$20 GeV is harder
than the electron spectrum and therefore, the positron fraction shows a striking increase
dubbed as {\em positron excess}~\cite{ref:pamela-pf,ref:ams-pf}.

The pure secondary production of positrons, originated from the interaction of primary cosmic
rays with the interstellar medium, can not account for the features in the observed spectrum.
Additional source terms for positron and electrons are needed to reproduce the measurements.

The physical origin of the source term has been extensively discussed. The proposed
models can be grouped into three different classes. First, secondary production in the standard
cosmic ray sources~\cite{ref:secacc}; second, exotic astrophysical sources, e.g.
pulsars~\cite{ref:pulsars}; finally, dark matter annihilation in the galactic halo~\cite{ref:galdm}
or in the Earth's vicinity~\cite{ref:sundm}. The detection of anisotropies in the positron arrival
directions would certainly favor some of these hypotheses.

Searches for large scale anisotropies in the positron samples have been carried out
by the experiments~\cite{ref:ams-iso,ref:pamela-iso}. The result is that the distribution of the
arrival directions is consistent with isotropy and thus, limits may be set
to the anisotropies at different angular scales. In particular, an upper limit to a dipole anisotropy
can be established within a certain confidence level.

In this note we present a short review of the mathematical rationale behind the analysis of dipole
anisotropies and the confidence regions one may achieve from an experiment. The basic definitions
are introduced in section~\ref{Basic Formulae}. In section~\ref{Parametric Inference}, the
expressions for the lower bound to the one-side confidence regions to a dipole anisotropy and
the expected limit from a single experiment are obtained. In section~\ref{Discussion}, we discuss the
consistency with the published limits by the AMS and PAMELA collaborations.

\section{Basic Formulae}
\label{Basic Formulae}

The angular distribution of cosmic rays is described by a probability
density function $p({\theta},{\phi}|{\bullet}){\in}L_2({\Omega})$, which is
a real, non-negative and well-behaved function with support in
${\Omega}=[0,{\pi}]{\times}[0,2{\pi}]$. The analysis of anisotropies
is based on the expansion of this density in a spherical harmonics basis.
Being a real function, it is natural to use the orthonormal real harmonics
basis (therefore real coefficients), though irrelevant for what follows.
Thus, after normalization, we can write
\begin{eqnarray}
\label{eqn:pdf}
p({\theta},{\phi}|\mbox{{\boldmath ${a}$}})\,=\,
\frac{\textstyle 1}{\textstyle 4{\pi}}\,
\left(1\,+\,a_{lm}\,Y_{lm}({\theta},{\phi})\right) 
\end{eqnarray}
where $\mbox{{\boldmath ${a}$}}=\{a_{lm};\,l{\geq}1,-l{\leq}m{\leq}l\}$,
\begin{eqnarray}
a_{lm}\,=\,
4{\pi}\int_{\Omega}p({\theta},{\phi})\,Y_{lm}({\theta},{\phi})\,d{\mu}
\,=\,4{\pi}\,E_{p;{\mu}}[Y_{lm}({\theta},{\phi})]\,,
\end{eqnarray}
$d{\mu}=\sin{\theta}d{\theta}d{\phi}$
and summation over repeated indices is understood. In this note,
we shall be particularly interested in dipole anisotropies $(l=1)$ so, to
simplify the notation, we redefine the indices 
$(l,m)=\{(1,-1),(1,0),(1,1)\}$ as $i=\{1,2,3\}$ and write
\begin{eqnarray}
\label{eqn:pthph}
p({\theta},{\phi}|\mbox{{\boldmath ${a}$}})
\,=\,\frac{1}{4{\pi}}\,\left( 1\,+\,
a_1Y_1\,+\,a_2Y_2\,+\,a_3Y_3 \right)
\end{eqnarray} 
with $\mbox{{\boldmath ${a}$}}=(a_1,a_2,a_3)$.

The coefficient of dipole anisotropy 
\begin{eqnarray}
{\delta}\,=\,\sqrt{\frac{3}{4{\pi}}}\,
\left( a_1^2\,+\,a_2^2\,+\,a_3^2 \right)^{1/2}
\end{eqnarray}
corresponds to the usual definition
~\cite{ref:pamela-pf,ref:ams-pf}
based on the maximum and minimum values of the intensity of Cosmic Rays.

Obviously, for any
$({\theta},{\phi}){\in}{\Omega}$ we have that
$p({\theta},{\phi}|\mbox{{\boldmath ${a}$}})
{\geq}0$ so the set of parameters $\mbox{{\boldmath ${a}$}}$ are constrained 
on a compact support. 
In the present case, non-negativity of the probability density
implies that the dipole coefficients are bounded by the sphere
\begin{eqnarray}
a_1^2\,+\,a_2^2\,+\,a_3^2\,{\leq}\,\frac{4{\pi}}{3}
\end{eqnarray}
and therefore, the coefficient of dipole anisotropy by
${\delta}\,{\leq}\,1$.

\section{Parametric Inference}
\label{Parametric Inference}

Given an exchangeable sequence $\{x_1,x_1,{\ldots},x_n\}$ of $n$
observations $x_i=({\theta}_i,{\phi}_i)$ identically distributed from
the parametric model 
$p({\theta},{\phi}|\mbox{{\boldmath ${a}$}})$, we shall be interested in
making inferences on the parameters 
$\mbox{{\boldmath ${a}$}}=(a_1,a_2,a_3)$ 
from the proper posterior density,
\begin{eqnarray}
p(\mbox{{\boldmath ${a}$}}|{\bullet})\,{\propto}\,
\pi(\mbox{{\boldmath ${a}$}})\,
\prod_{i=1}^n\,p({\theta}_i,{\phi}_i|\mbox{{\boldmath ${a}$}})
\end{eqnarray}
with $\pi(\mbox{{\boldmath ${a}$}})$ the reference prior
and, eventually,
on the dipole anisotropy coefficient $\delta$ 
from $p({\delta}|{\rm data})$.

\subsection{Expected precision on the dipole coefficients}
\label{Expected precision}
The first question to address is the precision we expect
on the parameters $\mbox{{\boldmath ${a}$}}$ given a sequence of observations.
The space of parameters 
${S}_{\mbox{{\boldmath ${a}$}}}$
is a Riemannian manifold with the Fisher-Rao metric tensor given by:
\begin{eqnarray}
g_{ij}( \mbox{{\boldmath ${a}$}})\,=\,
\frac{1}{(4{\pi})^2}\,
\int_{\Omega}\,
\frac{Y_i({\theta},{\phi})\,Y_j({\theta},{\phi})}
{p({\theta},{\phi}|\mbox{{\boldmath ${a}$}})}\,d{\mu}\,.
\end{eqnarray}
We shall be interested in a neighborhood of
$\mbox{{\boldmath ${a}$}}=(0,0,0)$ so,
expanding this expression 
to order ${\sl O}({a}_i^4)$ one gets:
\begin{eqnarray}
g_{ij}\,{\simeq}\,\frac{1}{4{\pi}}(1\,+\,\frac{{\delta}^2}{5})\,{\delta}_{ij}
\,+\,\frac{6}{5(4{\pi})^2}\,{a}_i{a}_j\,. 
\end{eqnarray}
Regularity conditions are satisfied to ensure that, for large $n$, 
the posterior density 
$p(\mbox{{\boldmath ${a}$}}|{\bullet})$
is asymptotically normal. Then, for large $n$, prior precision
can be ignored compared to that from data and the accuracy will be given
by the Hessian matrix. Thus, again to order ${\sl O}({a}_i^4)$,
we have that:
\begin{eqnarray}
V({a}_i,{a}_i)\,\simeq\,
\frac{4{\pi}}{n}\,\left(
1\,-\,\frac{3}{5(4{\pi})}
\left( {a}_1^2\,+\,{a}_2^2\,+\,{a}_3^2\,+\,
2\,{a}_i^2 \right) \right)
\end{eqnarray}
and covariances
\begin{eqnarray}
V({a}_i,{a}_j)\,\simeq\,
-\frac{6}{5\,n}\,{a}_i{a}_j
\hspace{1.5cm}i{\neq}j.
\end{eqnarray}
In particular, in the limit ${a}_{1,2,3}{\rightarrow}0$ one has
\begin{eqnarray}
{\sigma}({a}_i)\,=\,
\sqrt{\frac{4{\pi}}{n}}
\hspace{1.0cm}{\rm and}\hspace{1.0cm}{\rho}_{ij}\,=\,0\,.
\end{eqnarray}

A question that has been posed is whether one can have more precise
estimates. 
First, one can obviously rotate the
reference frame to increase the sensitivity in one particular direction.
However, on the one hand, if the coefficients $a_i$  have
small absolute value the
changes in the Hessian matrix will be of order
${\sl O}({a}_i^2)$. On the other hand,
rotations are orthogonal transformations, 
isometric, so 
for a second order tensor $G$
there are three quantities that remain
invariant (Cayley-Hamilton) 
and they ensure that, to order ${\sl O}({a}_i^2)$,
$\sum {\sigma}^2(a_i)$ and $\prod {\sigma}^2(a_i)$ will not change.
In consequence, any improvement in the accuracy of one parameter 
has to be compensated by the worsening of the accuracy
of the other parameters and therefore
there is no way to improve the precision of the coefficient of 
dipole anisotropy in a significant manner by rotations of the reference frame.

In a more realistic situation, two effects have to be considered. First, it may happen
that only a fraction of ${\Omega}$ is experimentally observed. 
This can be easily accounted for with the appropriate normalization of Eqn.~\ref{eqn:pdf}.
Second, it may happen that the observation time is not uniform over ${\Omega}$. 
This effect can be accounted for modifying the  
probability density (\ref{eqn:pdf}) as
\begin{eqnarray}
p^{\star}({\theta},{\phi}|\mbox{{\boldmath ${a}$}})
\,{\propto}\,f({\theta},{\phi})\,\left( 1\,+\,
a_1Y_1\,+\,a_2Y_2\,+\,a_3Y_3 \right)\,.
\end{eqnarray} 
with $f({\theta},{\phi})$ a non-negative function. In the first place, it should be noted that
if the coefficients $|a_i|$ are small, only the dipole and quadrupole terms of the
expansion of $f({\theta},{\phi})$ are relevant in the corresponding metric tensor. 
In the second place, even though the accuracy along a particular direction may improve it
can be shown (Schur's inequality)
that, under the hypothesis of no true anisotropy,
the quadratic sum of uncertainties is bounded from below by that obtained in the case of
uniform exposure and, in consequence, the limits obtained in the next section still apply.

\subsection{One-side credible regions for $\delta$}
\label{One-side credible regions}
\begin{figure}[!ht]
\begin{center}
\includegraphics[width=0.72\textwidth]{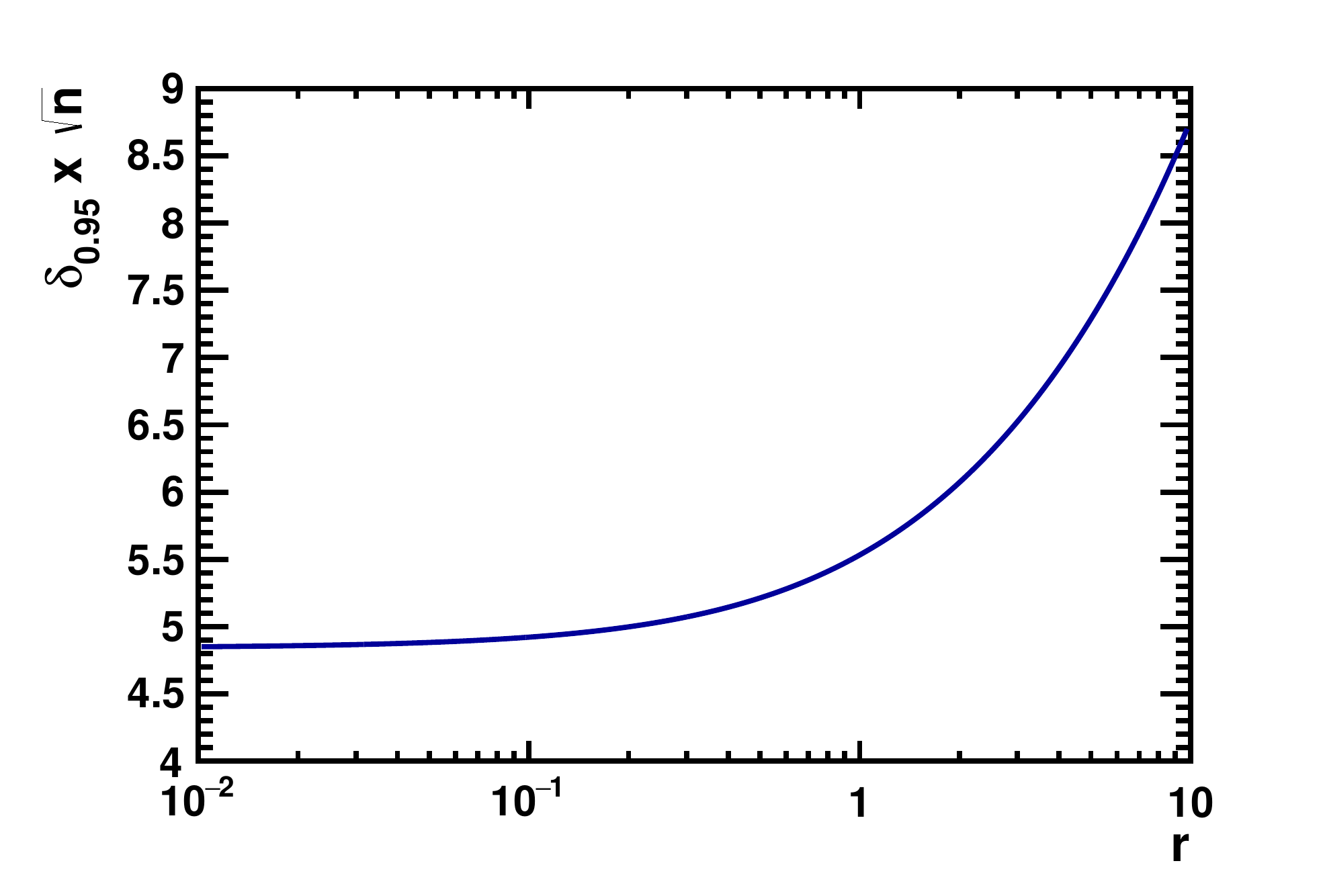}
\caption{Upper limit at 95\% CL on the dipole anisotropy parameter $\delta$ as a function of the
         measured parameter $r$ as defined in the text. $\delta_{0.95}$ is scaled with the
         factor $\sqrt{n}$, being $n$ the size of the sample.}
\label{fig:delta95_vs_lambda}
\end{center}
\end{figure}

In this problem, there are no sufficient minimal statistics 
neither for the coefficients $\mbox{{\boldmath ${a}$}}$ nor for
${\delta}$ other
than the whole sample. However,
as argued above, for large statistics and sufficiently smooth priors
we may consider that the 
posterior density 
is well approximated by
\begin{eqnarray}
p(\mbox{{\boldmath ${a}$}}|{\bullet})\,{\simeq}\,
\prod_{k=1}^{3}\,N({a}_k|m_k,{\sigma}_k^2)
\end{eqnarray}
The distribution of the coefficient of anisotropy ${\delta}$
is a quite involved expression. Nevertheless,
if we define the statistics
\begin{eqnarray}
{\sigma}_s^2\,=\,\frac{1}{3}
\sum_{k=1}^{3}{\sigma}_k^2\,;\hspace{1.cm}
s\,=\,
\frac{4{\pi}}{3{\sigma}_s^2}
\hspace{1.cm}{\rm and}\hspace{1.cm}
r\,=\,\sum_{k=1}^{3}
\left(\frac{m_k}{{\sigma}_k}\right)^2
\nonumber
\end{eqnarray}
and assume ${\sigma}_k^2{\simeq}{\sigma}_s^2;\,k=1,2,3$ 
we have the simpler expression
	\begin{eqnarray}
	  p({\delta}|r,s)\,=\,
	  \frac{s\sqrt{2}}{\sqrt{{\pi}r}}\,e^{-r/2}\,
	  e^{\textstyle -s {\delta}^2/2}\,{\delta}\,{\rm sinh}(\sqrt{rs}\,{\delta})\,
	  \mbox{{\boldmath $1$}}_{[0,{\infty})}({\delta})
	\end{eqnarray}
that, within the assumptions made, is the  
{\sl posterior density} 
for the anisotropy coefficient from which inferences should be drawn.
The first two moments are given by:
  \begin{eqnarray}
    E[{\delta}]\,=\,e^{-r/2}\frac{\sqrt{2}}{\sqrt{{\pi}s}}\,+\,
    \frac{1+r}{\sqrt{sr}}\,
         {\rm erf}\left[\sqrt{\frac{r}{2}}\right] 
	 \hspace{1.0cm}{\rm and}\hspace{1.0cm}
	 E[{\delta}^2]\,=\,\frac{3+r}{s}
\end{eqnarray}
and, after integration, one gets the distribution function:
	\begin{eqnarray}
	  P({\delta}{\leq}{\delta}_0|r,s)\,=\,
	  \frac{1}{2}\left(
	       {\rm erf}(w_+)+
	       {\rm erf}(w_-)
	       \right)
	       +\frac{1}{\sqrt{2{\pi}r}}\,
	       \left(
	       e^{-w_+^2}-e^{-w_-^2}
	       \right)
	\end{eqnarray}
where $\sqrt{2}\,w_{\pm}=({\delta}_0\sqrt{s}{\pm}\sqrt{r})$ that
can be used to obtain confidence regions. In particular, the value ${\delta}_{0.95}$
for the one-side 95\% credible region is given by
$P({\delta}{\leq}{\delta}_{0.95}|r,s)=0.95$
and its dependence with
$r$ is shown in Fig.~\ref{fig:delta95_vs_lambda} 
assuming that ${\sigma}^2_k=4{\pi}/n;\, k=1,2,3$.

In the limit $r{\rightarrow}0$ one gets a Generalized Gamma
Distribution:
\begin{eqnarray}
p({\delta}|s)\,=\,s^{3/2}\,
\sqrt{\frac{2}{\pi}}\,
e^{\textstyle -s {\delta}^2/2}\,{\delta}^2\hspace{1.cm}
\end{eqnarray}
with $E[{\delta}]=2\sqrt{2}/\sqrt{{\pi}s}$, 
$E[{\delta}^2]=3/s$ and the distribution function
\begin{eqnarray}
P({\delta}{\leq}{\delta}_0|s)\,=\,
\frac{2}{\sqrt{\pi}}\,
{\gamma}(\frac{3}{2},\frac{s{\delta}_0^2}{2})
\end{eqnarray}
with ${\gamma}(s,x)$ the Incomplete Gamma Function.

Assuming ${\sigma}^2_k=4{\pi}/n;\, k=1,2,3$
we have, for
$P({\delta}{\leq}{\delta}_{\alpha}|s)={\alpha}$, that
\begin{eqnarray}
\gamma(\frac{\textstyle 3}{\textstyle 2},
       \frac{\textstyle {\delta}_{\alpha}^2\,n}{\textstyle 6})
\,=\,{\alpha}\,\frac{\sqrt{\pi}}{2}
\end{eqnarray}
and, in particular, a 95\% one-side credible region:
\begin{eqnarray}
\label{eqn:CL95}
{\delta}^{0}_{0.95}\,=\,
\frac{4.84}{\sqrt{n}}
\end{eqnarray}
Within the assumptions and approximations made, this is the 
{\sl ``smallest''} 95\% upper limit 
one may set on $\delta$ from a sample of size $n$. 
Moreover, from the arguments given in the previous 
section it is clear that, in the case of no true anisotropy,  
collecting data in a subset of $\Omega$ or 
having a non-uniform exposure time can only increase this value.

Obviously, even in the ideal case of Eqn.~\ref{eqn:pthph} being the true underlying distribution, 
this is not necessarily the limit one will get from a particular realization of the
experiment. To have a feeling of what you expect for a sample of size $n$,
we can evaluate $E_r[{\delta}_{\alpha}(r,s)]$. For
${\alpha}=0.95$ we get, under the hypothesis of
no true anisotropies:
\begin{eqnarray}
\label{eqn:Mean}
{\delta}^{exp}_{0.95}\,=\,
\frac{6.39}{\sqrt{n}}
\end{eqnarray}

\begin{figure}[!ht]
\begin{center}
\includegraphics[width=0.72\textwidth]{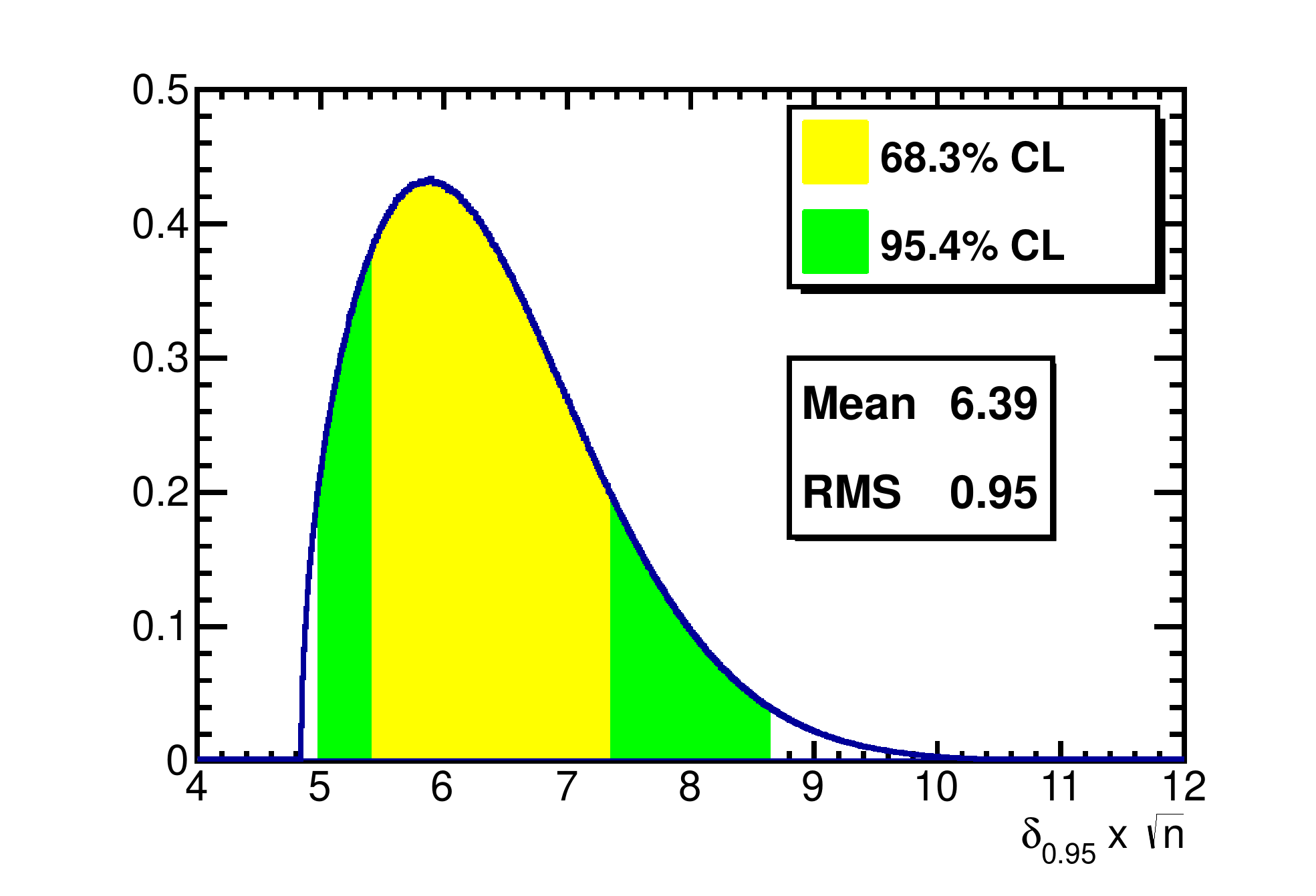}
\caption{Distribution of the upper limit at 95\% CL on the dipole anisotropy parameter
         $\delta$ expected for independent samples under the assumption of no true
         anisotropy. 
         The 68.3\% and 95.4\% equal tail area probability regions are displayed in
         yellow and green respectively.
         $\delta_{0.95}$ is scaled with the factor $\sqrt{n}$, being $n$ the size of the sample.}
\label{fig:delta95}
\end{center}
\end{figure}

The probability density function of $\delta_{0.95}$ is displayed in 
Fig.~\ref{fig:delta95} together with the 
68.3\% and 95.4\% equal tail area probability regions.

\section{Discussion}
\label{Discussion}

The conclusion from previous sections is that, under fairly general conditions, there is an absolute
lower bound to the one-side confidence regions for the coefficient of anisotropy determined from
a sample of events (Eqn.~\ref{eqn:CL95}). On the other hand, the limit one may expect from a single
experiment, under the assumption of no true anisotropy is given by Eqn.~\ref{eqn:Mean}.

We have checked the consistency of these results with the published limits by the AMS and PAMELA
collaborations.

\begin{figure}[!ht]
\begin{center}
\includegraphics[width=0.72\textwidth]{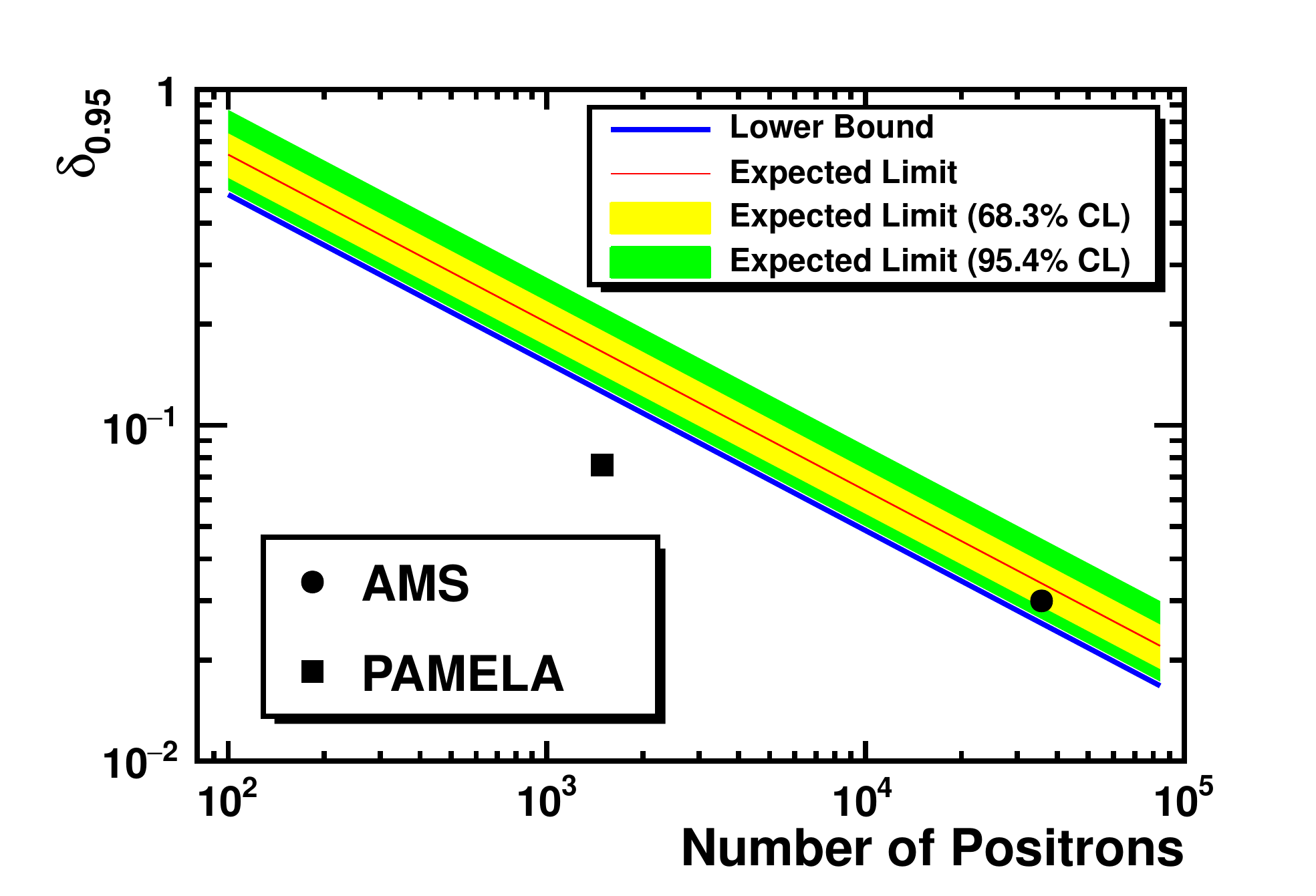}
\caption{Upper limit at 95\% CL on the dipole anisotropy parameter $\delta$ as a function of the
         number of detected positrons. The absolute lower bound (Eqn.~\ref{eqn:CL95}) is displayed
         in blue. The red curve, corresponds to the expected limit for a single experiment (Eqn.~\ref{eqn:Mean}).
         The equal tail probability bands containing the ensemble of 68.3\% and 95.4\% of the expected experimental
         limits are displayed in yellow and green respectively. The limit obtained by AMS~\cite{ref:ams-iso}
         is shown with a circle. The result from PAMELA~\cite{ref:pamela-iso} is displayed with a square.}
\label{fig:pamela_ams}
\end{center}
\end{figure}

In Fig.~\ref{fig:pamela_ams}, we display, as a function of the sample size, the absolute
lower bound for the 95\% CL exclusion limits, as well as the expected limit for a single experiment
and the equal tail probability 
bands containing the ensemble of 68.3\% and 95.4\% of them
under the assumption of no true anisotropy and ${\sigma}_k=4{\pi}/n;\,k=1,2,3$. 
AMS and PAMELA results are also displayed.

AMS obtained upper limits at the 95\% confidence level for cumulative energy ranges on the positron
over electron ratio~\cite{ref:ams-iso}. The limit obtained for the energy range from 16 to 350 GeV
on a sample of 35,000 positrons is ${\delta}_{0.95} = 0.030$, in good 
agreement with what one may expect under the assumption of no true anisotropy 
${\delta}^{exp}_{0.95}=0.034$ (Eqn.~\ref{eqn:Mean}) 
and above the limit ${\delta}^0_{0.95}=0.026$ given by Eqn.~\ref{eqn:CL95}.

PAMELA has recently released the search for anisotropies in cosmic-ray positrons~\cite{ref:pamela-iso}.
A dipole anisotropy upper limit of ${\delta}_{0.95} = 0.076$ at the 95\% confidence level is determined
from a data sample of 1,500 positrons selected in the rigidity range from 10 to 200 GV.
This value is not consistent with the lower bound given by Eqn.~\ref{eqn:CL95} for this sample size 
(${\delta}^0_{0.95}=0.125$) and, therefore, we suspect that
the upper bounds on the dipole anisotropy coefficient have not been derived correctly.



\end{document}